\documentclass[twocolumn]{article}
\pdfmapfile{+pikfonts.map}
\usepackage{amsmath,amssymb,bbm}
\usepackage[sflevel=full,color=full,addlogo=GaNe_logo_noborder]{pik}
\usepackage{bm}
\usepackage{cuted}

\shorttitle{%
Probability-Based Estimation}
\title{%
Probability-Based Estimation}
\date{%
This version \today
}
\shortauthor{%
Jobst Heitzig
}
\author{%
Jobst Heitzig
}
\affil{%
Potsdam Institute for Climate Impact Research, Complexity Science\\
FutureLab on Game Theory and Networks of Interacting Agents\\ 
P.\,O.\ Box 60 12 03, 14412 Potsdam, Germany\\
{\tt heitzig@pik-potsdam.de}
}

\def\ge{\geqslant}

\def\le{\leqslant}

\def\epsilon{\varepsilon}
\def\rho{\varrho}
\def\phi{\varphi}

\begin{document}

%%%%%%%%%%%%
% Abstract:
\twocolumn[\maketitle\begin{abstract}
We develop a theory of estimation when in addition to a sample of $n$ observed outcomes the underlying probabilities of the observed outcomes are known, as is typically the case in the context of numerical simulation modeling, e.g. in epidemiology.

For this enriched information framework, we design unbiased and consistent ``probability-based'' estimators whose variance vanish exponentially fast as $n\to\infty$, as compared to the power-law decline of classical estimators' variance.
\end{abstract}\vskip 4ex]

%%%%%%%%%%%%%
% Main text:

\section{Problem statement}

There is a discrete probability space with finite outcome set 
$\Omega$ and probability weight function $p:\Omega\to[0,1],\omega\mapsto p(\omega)$, $\sum_{\omega\in\Omega}p(\omega)=1$.
There is also an event $A\subseteq \Omega$ the probability of which, 
$\pi = \sum_{\omega\in A}p(\omega)$, we want to estimate.
We don't know $p$ but we do know $\Omega$ and $A$, in particular we know the number $m=|A|$ of outcomes in $A$.

We have access to a sampler which draws iid samples $\omega_1,\dots,\omega_n$ from $(\Omega, p)$ and which in addition (!) gives us the corresponding probabilities $x_1 = p(\omega_1),\dots,x_n = p(\omega_n)$.

{\em How to ``best'' make use of this additional information? E.g., what consistent (and maybe also unbiased) estimator of $\pi$ has the smallest standard error given this information?}

\subsection*{Use case: costly simulations} 

In an important class of use cases in which this occurs, each $\omega$ is a possible trajectory of some stochastic dynamical system that we can simulate, and the simulator allows us to compute $p(\omega)$ iteratively by multiplying up the probabilities of the changes performed in individual time steps. $A$ encodes some macroscopic event that we are interested in, such as: the system tips, an epidemic gets detected, the system converges back to a certain attractor, etc.

\subsubsection*{Application: Epidemic spreading on a network}

Assume a network (graph) $G=(V,E)$ and an SI infection process where initially all nodes are susceptible, at discrete time $t\ge 0$ node $v\in V$ has a basic probability of getting infected of $p_1(v,t)$, and independently for each edge $e=\{v,v'\}\in E$ with infected $v$, $v'$ has a transmission probability of getting infected of $p_2(v,v',t)$ (e.g., \cite{ansari2021temporal}.
Finally, there is a sequence $((s_j,\tau_j))_j$ with sentinel nodes $s_j\in V$ and testing time points $\tau_j\le T$.
The event $A$ is the fact that an outbreak has been detected by one of the latter tests, i.e., for at least one $j$, $s_j$ is infected at time $\tau_j$.
If the network is complex, there is no simple analytical solution for $P(A)$, hence we assume the SI process has been simulated $n$ times from $t=1$ to $t=T$ and $\omega_i=(\omega_{ivt})_{v\in V,t\in\{1,\dots,T\}}$ is the binary matrix encoding whether each node $v$ was infected at each time $t$.
As the simulator can easily track the probability $x_i$ of each realized trajectory $\omega_i$, this information can be used in estimating $\pi$.

\paragraph{Toy example.}

As a simple analytically tractable example assume $G$ is a chain of $L+1$ nodes $v=0\dots L$, $p_1(0,0) = p_1 > 0$, $p_1(v,t) = 0$ for all other $v,t$, $p_2(v,v',t) \equiv p_2 > 0$, and there is only one test at $s_1=L$ at time $\tau_1=T$.
Then $A$ is the event that node $L$ is infected at time $T$. 
The only $\omega$ that have positive probability are those where for each infected node $v$ at $t$, all $v'\le v$ are infected at $t$, $v$ remains infected at all $t'\ge t$, and either $v$ was already infected at $t-1$, or $v+1$ is not yet infected at $t$.
Let us encode such an $\omega$ by the tuple of time points $t_1<\cdots<t_L$ at which nodes $1\dots L$ get first infected, where $t_v\in[v,\infty]$.
With $q_2=1-p_2$, the probability of this $\omega$ is 
\begin{align}
    p(\omega) &= p(t_1,\dots,t_L) \\
    &= p(t_1,\dots,t_{L-1}) q_2^{t_L-t_{L-1}-1} p_2 \\
    &= p_1 q_2^{t_L-L} p_2^L.
\end{align}
The event $A$ corresponds to $t_L\le T$ and has thus probability
\begin{align}
    \pi &= \sum_{1\le t_1<\cdots<t_L\le T} p_1 q_2^{t_L-L} p_2^L = p_1 (p_2/q_2)^L \sum_{t_L=L}^T {t_L-1\choose L-1} q_2^{t_L} \\
    &= p_1 \left(1 - p_2^L q_2^{T+1-L}{T\choose L-1}\;{}_2F_1(1,T+1;T+2-L;q_2)\right),
\end{align}
where ${}_2F_1$ is the hypergeometric function. 
As we can see, this is already a rather complicated formula even for this simplest case of a network and just one test.
Later we will also need the fact that the opposite event $\neg A$ has probability
\begin{align}
    1 - \pi &= 1 - p_1 + \sum_{1\le t_1<\cdots<t_L>T} p_1 q_2^{t_L-L} p_2^L \\
    &= 1 - p_1 + p_1 (p_2/q_2)^L \sum_{t_L=T+1}^\infty {t_L-1\choose L-1} q_2^{t_L}.
\end{align}

\section{Benchmark: relative frequency}

As is well-known, without knowledge of the probabilities $x_i$, the most straightforward estimator $\hat\pi_0$ of $\pi$ is the relative frequency
\begin{align}
    \hat\pi_0 &= |\{i:\omega_i\in A\}| / n.
\end{align}
That estimator is unbiased, consistent, and has variance
\begin{align}
    v_0 &= \pi (1 - \pi) / n,
\end{align}
which can be estimated by the plug-in estimator
\begin{align}
    \hat v_0 &= \hat\pi_0 (1 - \hat\pi_0) / n.
\end{align}
Since the estimator is unbiased, its standard error is simply $\sqrt{v_0} = \sqrt{\pi (1 - \pi) / n}$, a very well-known fact.

Any estimator using also the additional information given by the $x_i$ must be compared against this benchmark.

An obvious improvement is to use 
\begin{align}
    \hat \pi_{0,\max{}} &= \max(\hat\pi_0, \sum_{\omega\in O\cap A} p(\omega)),
\end{align}
where $O = \{\omega_1,\dots,\omega_n\}$ is the set of observed outcomes. This clearly has a smaller standard error (if only negligibly smaller), but it is {\em not unbiased} and surely not optimal in any sense yet. 

\section{Idea 1: use a weighted sum of the observed probabilities}

Let 
\begin{align}
    q(\omega) & = 1-p(\omega), \\
    O &= \{\omega_1,\dots,\omega_n\},
\end{align}
the latter being the set of observed outcomes (counting each distinct outcome only once!), and note that we know $p(\omega)$ for each $\omega\in O$ (it equals one of the $x_i$).
Then the following is a consistent and unbiased estimator of $\pi$:
\begin{align}
    \hat\pi_1 = \sum_{\omega\in O\cap A}\frac{p(\omega)}{1 - q(\omega)^n}.
\end{align}
It is consistent because for $n\to\infty$, $O\to\Omega$ almost surely, and $[1-p(\omega)]^n\to 0$ for all $\omega$ with $p(\omega)>0$.
It is unbiased because 
\begin{align}
    \mathbb{E}\hat\pi_1 &= \sum_{\omega\in A}\frac{p(\omega)}{1 - q(\omega)^n}\mathbb{E}\mathbbm{1}_O(\omega) \\
    &= \sum_{\omega\in A}\frac{p(\omega)}{1 - q(\omega)^n}(1 - q(\omega)^n) \\
    &= \sum_{\omega\in A}p(\omega) = \pi,
\end{align}
where $\mathbbm{1}_O$ is the indicator function of $O$
and $1 - q(\omega)^n$ is the probability that $\omega\in O$.
What is its standard error?
We have 
\begin{align}
    \mathbb{E}\hat\pi_1^2 &= \sum_{\omega,\omega'\in A}\frac{p(\omega)}{1 - q(\omega)^n}\frac{p(\omega')}{1 - q(\omega')^n}\mathbb{E}(\mathbbm{1}_{\omega\in O}\mathbbm{1}_{\omega'\in O}) \\
    &= \sum_{\omega,\omega'\in A,~\omega\neq\omega'}\frac{p(\omega)}{1 - q(\omega)^n}\frac{p(\omega')}{1 - q(\omega')^n}\times \nonumber\\
    &\quad\quad\times (1 - q(\omega)^n - q(\omega')^n + [1-p(\omega)-p(\omega')]^n) \nonumber\\
    &\quad + \sum_{\omega\in A}\frac{p(\omega)^2}{(1 - q(\omega)^n)^2}(1 - q(\omega)^n) \\
    &= \sum_{\omega,\omega'\in A}\frac{p(\omega)}{1 - q(\omega)^n}\frac{p(\omega')}{1 - q(\omega')^n}\times \nonumber\\
    &\quad\quad\times (1 - q(\omega)^n - q(\omega')^n + [1-p(\omega)-p(\omega')]^n) \nonumber\\
    &\quad + \sum_{\omega\in A}\frac{p(\omega)^2}{(1 - q(\omega)^n)^2}(q(\omega)^n - [1-2p(\omega)]^n).
\end{align}
(The final bracket in the second line equals $1-P(\omega\notin O) - P(\omega'\notin O)+P(\omega,\omega'\notin O)$).
The exact variance of $\hat\pi_1$ is then
\begin{align}
    v_1 &= \mathbb{E}\hat\pi_1^2 - \pi^2 \\
    &= \sum_{\omega,\omega'\in A}p(\omega)p(\omega')\times \nonumber\\
    &\quad\quad\times \left[\frac{1 - q(\omega)^n - q(\omega')^n + [1-p(\omega)-p(\omega')]^n}{(1 - q(\omega)^n)(1 - q(\omega')^n)} - 1\right] \nonumber\\
    &\quad + \sum_{\omega\in A}p(\omega)^2\frac{q(\omega)^n - [1-2p(\omega)]^n}{(1 - q(\omega)^n)^2},
\end{align}

\paragraph{Toy example.}
In our toy example from the introduction, a numerical estimation of $v_0$ and $v_1$ shows that for $L=10$, $T=20$, already at $n=10$ we have $v_1<v_0$, improving fast as $n$ grows. 

\paragraph{Asymptotic variance.}
For large $n$, we have 
\begin{align}
    v_1 &\approx \sum_{\omega,\omega'\in A}p(\omega)p(\omega')[1-p(\omega)-p(\omega')]^n 
    + \sum_{\omega\in A}p(\omega)^2 q(\omega)^n \\
    &\le m^2\bar p^2(1-2\underline p)^n 
    + m \bar p^2 (1-\underline p)^n \sim m \bar p^2 (1-\underline p)^n ,
\end{align}
where $\underline{p}=\min_{\omega\in A}p(\omega)$ and $\bar p=\max_{\omega\in A}p(\omega)$.
This bound declines exponentially fast with $n$ rather than just as an $O(1/n)$ like for the relative frequency!

In other words, asymptotically for $n\to\infty$, $\hat\pi_1$ will vastly outperform $\hat\pi^0$, but we don't know when that asymptotics kicks in. 
For large $\Omega$, it seems likely that a very large $n$ will be needed for $\hat\pi_1$ to outperform $\hat\pi^0$.

%Can we find a lower bound on $v_1$ to verify this?

\paragraph{Dependence of variance on distribution.}
If the probability mass within $A$ is distributed equally among $m$ different $\omega$, then 
\begin{align}
    v_1 &= \pi^2\left[\frac{1 - 2[1-\pi/m]^n + [1-2\pi/m]^n}{(1 - [1-\pi/m]^n)^2} - 1\right]+\nonumber\\
    &\quad  + \pi^2\frac{[1-\pi/m]^n - [1-2\pi/m]^n}{m(1 - [1-\pi/m]^n)^2} \\
    &= \pi^2\frac{
        (m-1)[1-2\pi/m]^n
        - m[1-\pi/m]^{2n}
        + [1-\pi/m]^n
    }{m(1 - [1-\pi/m]^n)^2}.
\end{align}
% Note that for small $xn$, we have $(1-x)^n \approx 1 - xn + x^2n(n-1)/2$. 
% So if $n\pi/m$ is small, we have
% \begin{align}
%     v_1 &\approx \frac{
%         \pi(1 - \pi - (n-1)\pi/2m)
%     }{n} < \frac{
%         \pi(1 - \pi)
%     }{n} = v_0.
% \end{align}
% In other words, the more concentrated the probability mass inside $A$, the more advantage $\pi_1$ has compared to $\pi_0$. More generally, we expect $\pi_1$ to be the better the more diverse the probabilities of individual $\omega\in A$ are.

\paragraph{Variance estimation.} The quantity
\begin{align}
    \sum_{\omega\in A} p(\omega)^2 q(\omega)^n
\end{align}
occurring in the above approximation can be estimated without bias by
\begin{align}
    \sum_{\omega\in A\cap O} \frac{p(\omega)^2 q(\omega)^n}{1 - q(\omega)^n}.
\end{align}
Similarly, $v_1$ can be estimated without bias by
\begin{align}
    \hat v_1 &= \sum_{\omega,\omega'\in A\cap O}p(\omega)p(\omega')\left[\frac{1}{(1 - q(\omega)^n)(1 - q(\omega')^n)} \right. \nonumber\\
    &\quad\quad\left.~ - \frac{1}{1 - q(\omega)^n - q(\omega')^n + [1-p(\omega)-p(\omega')]^n}\right] \nonumber\\
    &\quad + \sum_{\omega\in A\cap O}p(\omega)^2\frac{q(\omega)^n - [1-2p(\omega)]^n}{(1 - q(\omega)^n)^3}.
\end{align}

\paragraph{Dual and combined estimators.}
While $\hat\pi_1$ estimates $\pi$ based on $O\cap A$, one can of course also estimate $1-\pi$ based on $O-A$ in the same fashion. This gives another unbiased estimator of $\pi$:
\begin{align}
    \hat\pi'_1 = 1 - \sum_{\omega\in O- A}\frac{p(\omega)}{1 - q(\omega)^n}.
\end{align}
Now it seems that a suitable (convex) combination of $\hat\pi_0$, $\hat\pi_1$ and $\hat\pi'_1$ should still be unbiased and have even smaller variance. {\em But which combination is optimal?} If the three estimators were independent, the following convex combination would have minimal variance:
$(\hat\pi_0/v_0 + \hat\pi_1/v_1+ \hat\pi'_1/v'_1) / (1/v_0 + 1/v'_1 + 1/v_1)$.
Since we don't know $v_1,v'_1$, we can only use their estimates, leading to the estimator
\begin{align}
    \hat v_1'' &= \frac{\hat\pi_0/\hat v_0 + \hat\pi_1/\hat v_1+ \hat\pi'_1/\hat v'_1}{1/\hat v_0 + 1/\hat v'_1 + 1/\hat v_1}
\end{align}
(where $\hat v'_1$ is like $\hat v_1$ with $O-A$ in place of $A\cap O$). 

\subsection{Generalization to mean estimation}
If the goal is to estimate the expected value $\mu = \mathbb{E}X$ of an observable random variable $X:\Omega\to \mathbb{R}$ rather than the probability of an event, one can do
\begin{align}
    \hat\mu'_1 = \xi + \sum_{\omega\in O}\frac{p(\omega)(X(\omega) - \xi)}{1 - q(\omega)^n}
\end{align}
for any arbitrary reference point $\xi$, which still gives an unbiased estimate.

{\em What choice of $\xi$ minimizes the variance of $\hat\mu_1$?}
The variance is
\begin{align}
    v'_1 &= \mathbb{E}\hat\mu_1^2 - \mu^2 \nonumber\\
    &= \xi^2 + 2\xi (\mu - \xi)  - \mu^2 \nonumber\\
    &\quad + \sum_{\omega,\omega'}p(\omega)X(\omega)p(\omega')X(\omega')\times \nonumber\\
    &\quad\quad\times \frac{1 - q(\omega)^n - q(\omega')^n + [1-p(\omega)-p(\omega')]^n}{(1 - q(\omega)^n)(1 - q(\omega')^n)} \nonumber\\
    &\quad - 2\xi\sum_{\omega,\omega'}p(\omega)X(\omega)p(\omega')\times \nonumber\\
    &\quad\quad\times \frac{1 - q(\omega)^n - q(\omega')^n + [1-p(\omega)-p(\omega')]^n}{(1 - q(\omega)^n)(1 - q(\omega')^n)} \nonumber\\
    &\quad + \xi^2\sum_{\omega,\omega'}p(\omega)p(\omega')\frac{1 - q(\omega)^n - q(\omega')^n + [1-p(\omega)-p(\omega')]^n}{(1 - q(\omega)^n)(1 - q(\omega')^n)} \nonumber\\
    &\quad + \sum_{\omega}p(\omega)^2X(\omega)^2\frac{q(\omega)^n - [1-2p(\omega)]^n}{(1 - q(\omega)^n)^2} \nonumber\\
    &\quad - 2\xi\sum_{\omega}p(\omega)X(\omega)p(\omega')\frac{q(\omega)^n - [1-2p(\omega)]^n}{(1 - q(\omega)^n)^2} \nonumber\\
    &\quad + \xi^2\sum_{\omega}p(\omega)p(\omega')\frac{q(\omega)^n - [1-2p(\omega)]^n}{(1 - q(\omega)^n)^2}
\end{align}
and its derivative w.r.t.\ $\xi$ is
\begin{align}
    \partial_\xi v &= 2\mu - 2\xi \nonumber\\
    &\quad - 2\sum_{\omega,\omega'}p(\omega)X(\omega)p(\omega')\times \nonumber\\
    &\quad\quad\times \frac{1 - q(\omega)^n - q(\omega')^n + [1-p(\omega)-p(\omega')]^n}{(1 - q(\omega)^n)(1 - q(\omega')^n)} \nonumber\\
    &\quad + 2\xi\sum_{\omega,\omega'}p(\omega)p(\omega')\times \nonumber\\
    &\quad\quad\times \frac{1 - q(\omega)^n - q(\omega')^n + [1-p(\omega)-p(\omega')]^n}{(1 - q(\omega)^n)(1 - q(\omega')^n)} \nonumber\\
    &\quad - 2\sum_{\omega}p(\omega)X(\omega)p(\omega')\frac{q(\omega)^n - [1-2p(\omega)]^n}{(1 - q(\omega)^n)^2} \nonumber\\
    &\quad + 2\xi\sum_{\omega}p(\omega)p(\omega')\frac{q(\omega)^n - [1-2p(\omega)]^n}{(1 - q(\omega)^n)^2},
\end{align}
which is zero if
\begin{align}
    \xi &= \left[\mu - \sum_{\omega,\omega'}p(\omega)X(\omega)p(\omega')\times\right. \nonumber\\
    &\quad\quad\quad\quad \times \frac{1 - q(\omega)^n - q(\omega')^n + [1-p(\omega)-p(\omega')]^n}{(1 - q(\omega)^n)(1 - q(\omega')^n)} \nonumber\\
    &\quad\quad\left. - \sum_{\omega}p(\omega)X(\omega)p(\omega')\frac{q(\omega)^n - [1-2p(\omega)]^n}{(1 - q(\omega)^n)^2}\right] \nonumber\\
    &\quad \Bigg / \left[1 - \sum_{\omega,\omega'}p(\omega)p(\omega')\times\right. \nonumber\\
    &\quad\quad\quad\quad \times \frac{1 - q(\omega)^n - q(\omega')^n + [1-p(\omega)-p(\omega')]^n}{(1 - q(\omega)^n)(1 - q(\omega')^n)} \nonumber\\
    &\quad\quad\left.  - \sum_{\omega}p(\omega)p(\omega')\frac{q(\omega)^n - [1-2p(\omega)]^n}{(1 - q(\omega)^n)^2}\right].
\end{align}
For large $n$, this is approximately $\mu$. 
This implies that a good choice of $\xi$ is an independent estimate of $\mu$ such as the sample mean $\xi = \sum_i X(\omega_i)/n$.

Getting back to the original case of probability estimation, where $X$ is the indicator function $1_A$, we now see that a further improvement of $\hat\pi_1$ is
\begin{align}
    \hat\pi'_1 &= \hat\pi_0 + \sum_{\omega\in O}\frac{p(\omega)(1_A(\omega) - \hat\pi_0)}{1 - q(\omega)^n} \\
    &= \hat\pi_1 + \left[1 - \sum_{\omega\in O}\frac{p(\omega)}{1 - q(\omega)^n}\right]\hat\pi_0.
\end{align}

% (???).
%\begin{align}
%    \frac{\sum_{\omega}p(\omega)q(\omega)^n[\mu + X(\omega)]}{ 2\sum_{\omega}p(\omega)q(\omega)^n}
%\end{align}
%which is dominated by those $\omega$ with large $q(\omega)^n$, i.e., with small $p(\omega)$.

\section{Idea 2: estimate the mean outcome probability}

We note that $\pi = m \xi$ where $\xi = \sum_{\omega\in A}p(\omega) / m$ is the average probability of the outcomes in $A$.
Also, $\xi$ can be interpreted as the expected value of $p(\omega_i)$ when an $\omega\in A$ is drawn uniformly (!) at random (rather than with relative probabilities $p(\omega)$).
Each $x_i$ of an $i$ with $\omega_i\in A$ can be seen as an estimate of $\xi$.
W.l.o.g.\ let us order the sample so that $\omega_1,\dots,\omega_k\in A$ and $\omega_{k+1},\dots,\omega_n\notin A$.
Then also each weighted average $\sum_{i=1}^k w_i x_i$ of the $k$ values $x_1,\dots,x_k$, with $\sum_i w_i = 1$, is an estimate of $\xi$.

To make such an estimate unbiased, we need to choose the averaging weights $w_i$ taking account of the fact that the $\omega_i$ were {\em not} sampled uniformly from $A$ but using the distribution given by $p$.
The correct averaging weight $w_i$ for $x_i$ must thus be proportional to the ratio between the uniform probability $1 / m$ and the actually used probability $p(\omega_i) / \pi$. In other words, we need
$w_i \propto (1/m) / (p(\omega_i)/\pi) \propto 1/x_i$.
This results in the estimators
\begin{align}
    \hat\xi 
    &= \frac{\sum_{i=1}^k \frac{1}{x_i}x_i}{\sum_{i=1}^k \frac{1}{x_i}} 
    = \frac{k}{\sum_{i=1}^k \frac{1}{x_i}}, \\    
    \hat\pi_2 
    &= m \hat\xi = \frac{m k}{\sum_{i=1}^k \frac{1}{x_i}}.    
\end{align}
In other words, rather than using the arithmetic mean of the $x_i$ to estimate $\xi$, we use the harmonic mean.

Indeed, the expected value of $\hat\xi$ is
\begin{align}
    \mathbb{E}\hat\xi &= \sum_{\omega_1\dots\omega_n}\left(\prod_i p(\omega_i)\right)\frac{\sum_i 1_A(\omega_i)}{\sum_i \frac{1_A(\omega_i)}{p(\omega_i)}}
\end{align}

\paragraph{Variance.}
Because $\hat\xi$ is the harmonic mean of the $x_i$, which are an iid sample from the distribution given by $p'(\omega) = p(\omega)/\pi$ on $A$, it is unbiased and its variance $u$ is
\begin{align}
    u &= \frac{\theta^4 \sigma^2}{k} 
    = \frac{\pi^2}{m^4 k}\left[\sum_{\omega\in A} \frac{1}{p'(\omega)} - m^2\right],
\end{align}
where
\begin{align}
    \theta &= 1 / E_{p'}\left[\frac{1}{X}\right] = 1 / \sum_{\omega\in A} p'(\omega)\frac{1}{X(\omega)} = 1 / \sum_{\omega\in A} \frac{1}{\pi} = \frac{\pi}{m}, \\
    \sigma^2 &= E_{p'}\left[\frac{1}{X} - \frac{1}{\theta}\right]^2 = \sum_{\omega\in A} p'(\omega) \left[\frac{1}{p(\omega)} - \frac{m}{\pi}\right]^2 \\
    &= \frac{1}{\pi}\sum_{\omega\in A} p(\omega) \left[\frac{1}{p(\omega)^2} - \frac{2m}{\pi p(\omega)} + \frac{m^2}{\pi^2}\right] \\
    &= \frac{1}{\pi}\sum_{\omega\in A} \frac{1}{p(\omega)} - \frac{m^2}{\pi^2}.
\end{align} 
From the sampled $x_i$, this variance can be estimated using standard methods, e.g., using the jackknife (leave-one-out) method:
\begin{align}
    \hat u 
    &= \frac{k - 1}{k} \sum_{i=1}^k \left(\hat\xi - \frac{k - 1}{\sum_{j\neq i} \frac{1}{x_j}}\right)^2
    = \frac{k - 1}{k} \sum_{i=1}^k \left(1 - \frac{k - 1}{k - \frac{\hat\xi}{x_i}}\right)^2 \hat\xi^2. 
\end{align}
For large $k$, this is approximately 
\begin{align}
    \hat u 
    &\approx \frac{k - 1}{k^3} \sum_{i=1}^k \left(1 - \frac{\hat\xi}{x_i}\right)^2 \hat\xi^2. 
\end{align}
The variance of $\hat\pi_2$ is then
\begin{align}
    v_2 &= m^2 u = \frac{\pi^2}{m^2 k}\left[\sum_{\omega\in A} \frac{1}{p'(\omega)} - m^2\right]
\end{align}
which can be estimated as
\begin{align}
    \hat v_2 &= m^2 \hat u 
    = \frac{k - 1}{k} \sum_{i=1}^k \left(1 - \frac{k - 1}{k - \frac{\hat\xi}{x_i}}\right)^2 \hat\pi_2^2. 
\end{align}

\section{Generalization to importance sampling}

Assume now that the $\omega_i$ are not from the ``distribution of interest'' $p$ but some other ``sampling'' distribution $p'$, that both $x_i = p(\omega_i)$ and $y_i = p'(\omega_i)$ are known, and that still we want to estimate $\pi = \sum_{\omega\in A} p(\omega)$. 
Put $z_i = x_i/y_i$.

The relative frequency estimator of $\pi$ is then replaced by the standard estimator from importance sampling \cite{tokdar2010importance}, 
\begin{align}
    \hat\pi_0 &= \frac{\sum_{i=1}^k z_i}{\sum_{i=1}^n z_i},
\end{align}
for which we do not need to know the $x_i$ or the $y_i$ but only the $z_i$.

Put $q'(\omega)=1-p'(\omega)$.
Our novel estimators $\hat\pi_1$ and $\hat\pi_2$ should then be defined as
\begin{align}
    \hat\pi_1 &= \sum_{\omega\in O\cap A}\frac{p(\omega)}{1 - q'(\omega)^n}, \\
    \hat\pi_2 
    &= m\hat\xi, \quad
    \hat\xi = \frac{\sum_{i=1}^k \frac{1}{y_i}x_i}{\sum_{i=1}^k \frac{1}{y_i}} 
    = \frac{Z}{W},
    & Z &= \sum_{i=1}^k z_i,
    & W &= \sum_{i=1}^k \frac{1}{y_i},    
\end{align}
and their variance can be calculated or estimated as
\begin{align}
    v_1 &= \sum_{\omega,\omega'\in A}p(\omega)p(\omega')\times \nonumber\\
    &\quad\quad\times \left[\frac{1 - q'(\omega)^n - q'(\omega')^n + [1-p'(\omega)-p'(\omega')]^n}{(1 - q'(\omega)^n)(1 - q'(\omega')^n)} - 1\right] \nonumber\\
    &\quad + \sum_{\omega\in A}p(\omega)^2\frac{q'(\omega)^n - [1-2p'(\omega)]^n}{(1 - q'(\omega)^n)^2}, \\
    \hat v_1 &= \sum_{\omega,\omega'\in A\cap O}p(\omega)p(\omega')\left[\frac{1}{(1 - q'(\omega)^n)(1 - q'(\omega')^n)}\right. \nonumber\\
    &\quad\quad \left. - \frac{1}{1 - q'(\omega)^n - q'(\omega')^n + [1-p'(\omega)-p'(\omega')]^n}\right] \nonumber\\
    &\quad + \sum_{\omega\in A\cap O}p(\omega)^2\frac{q'(\omega)^n - [1-2p'(\omega)]^n}{(1 - q'(\omega)^n)^3}, \\
    \hat v_2 
    &= m^2\frac{k - 1}{k} \sum_{i=1}^k \left(\hat\xi - \frac{\sum_{j\neq i} z_j}{\sum_{j\neq i} \frac{1}{y_j}}\right)^2 \\
    &= m^2\frac{k - 1}{k} \sum_{i=1}^k \left(\frac{Z}{W} - \frac{Z - z_i}{W - \frac{1}{y_i}}\right)^2. 
\end{align}

As in the standard theory of importance sampling, one can now ask how the sampling distribution $p'$ should be chosen to minimize $v_1$ or $v_2$, assuming that one has some influence on the choice of $p'$.

For large $n$, we have roughly 
\begin{align}
    v_1 &\approx \sum_{\omega\in A} p(\omega)^2 q'(\omega)^n.
\end{align}
Let's see whether we can find the optimal $p'$ simply via first-order conditions.
Shifting an infinitesimal sampling probability mass $dp'$ from $p'(\omega)$ to $p'(\omega')$ changes this by 
\begin{align}
    dv_1 &\approx n p(\omega)^2 q'(\omega)^{n-1} - n p(\omega')^2 q'(\omega')^{n-1}.
\end{align}
Setting this to zero for all $\omega\in A$ would imply that $p(\omega)^2 q'(\omega)^{n-1}$ is constant, hence 
\begin{align}
    p'(\omega) &= 1 - C p(\omega)^{-2/(n-1)}   
\end{align}
for some constant $C$, hence
\begin{align}
    1 &= \sum_{\omega\in A} p'(\omega) = |A| - C\sum_{\omega\in A}p(\omega)^{-2/(n-1)}, \\
    C &= (|A| - 1) / \sum_{\omega\in A}p(\omega)^{-2/(n-1)}, \\
    p'(\omega) &= 1 - \frac{|A| - 1}{|A|}\frac{p(\omega)^{-2/(n-1)} }{\langle p(\omega')^{-2/(n-1)}\rangle_{\omega'\in A}},   
\end{align}
which might be smaller than 0. So the optimal $p'$ will likely be a boundary solution with some $p'(\omega)=0$ in general rather than an interior solution given by the above equation.
Ansatz: $p'(\omega)=0$ whenever $p(\omega)<\alpha$ for some $\alpha$, and 
\begin{align}
    p'(\omega) &= 1 - C p(\omega)^{-2/(n-1)}   
\end{align}
whenever $p(\omega)\ge\alpha$, hence
\begin{align}
    p'(\omega) &= 1 - \frac{|A'(\alpha)| - 1}{|A'(\alpha)|}\frac{p(\omega)^{-2/(n-1)} }{\langle p(\omega')^{-2/(n-1)}\rangle_{\omega'\in A'(\alpha)}},   
\end{align}
where $A'(\alpha) = \{\omega\in A:p(\omega)\ge\alpha\}$ and $\alpha$ is the smallest value for which all $p'(\omega)$ thus computed are non-negative. This is probably the smallest $\alpha$ for which
\begin{align}
    \frac{|A'(\alpha)| - 1}{|A'(\alpha)|}\frac{\alpha^{-2/(n-1)} }{\langle p(\omega')^{-2/(n-1)}\rangle_{\omega'\in A'(\alpha)}} &\le 1.
\end{align}
Because always $\alpha^{-2/(n-1)}  / \langle p(\omega')^{-2/(n-1)}\rangle_{\omega'\in A'(\alpha)} > 1$, the factor $(|A'(\alpha)| - 1) / |A'(\alpha)|$ needs to compensate for this to get the product $\le 1$, hence the resulting set $A'(\alpha)$ is likely small, i.e., only a few $\omega$ with the largest $p(\omega)$ get a positive $p'(\omega)$. Since for these largest $p(\omega)$, the values $p(\omega)^{-2/(n-1)}$ are all close to 1, the resulting $p'(\omega)$ are all approx. $1/|A'(\alpha)|$. In other words, selecting a suitable number of $\omega\in A$ with the largest $p(\omega)$ and then sampling uniformly from them is close to optimal.

\section{Application: hypothesis testing in epidemic control}

Assume now that we want to test the hypothesis $H_0$ that an epidemic outbreak of type SI has occurred in a population into which the respective disease is introduced from the outside with a known probability $p_1$ per time and individual and can be transmitted with a known probability $p_2$ whenever two individuals meet, and that we know the contact network and have performed a number of tests for infection at certain nodes and timepoints, all of which turned out negative.

We can then simulate $n$ potential outbreaks and corresponding sets of tests, giving trajectories $\omega_i$ and corresponding probabilities $x_i$, and observe which simulations resulted in all tests being negative, $\omega_i\in A$, and which resulted in at least one test being positive $\omega_i\in\Omega\setminus A$.

Using the above designed methods, one can then estimate the probability $\pi$ of all tests being negative under the hypothesis $H_0$ of an outbreak having occurred. If this probability is below the set level of the test, say $0.01$, one would then reject the hypothesis and conclude that no outbreak has occurred.

%%%%%%%%%%%%%%
% References:

\small\sffamily

\subsection*{Funding}
This work was supported by the German Bundesministerium f\"ur Bildung und Forschung, FKZ 01KI1812 as part of the Forschungsnetz Zoonosen.

\bibliographystyle{plain}
\bibliography{refs}

\begin{thebibliography}{1}

\bibitem{ansari2021temporal}
Sara Ansari, Jobst Heitzig, Laura Brzoska, Hartmut~HK Lentz, Jakob Mihatsch,
  J{\"o}rg Fritzemeier, and Mohammad~R Moosavi.
\newblock A temporal network model for livestock trade systems.
\newblock {\em Frontiers in Veterinary Science}, page 1438, 2021.

\bibitem{tokdar2010importance}
Surya~T Tokdar and Robert~E Kass.
\newblock Importance sampling: a review.
\newblock {\em Wiley Interdisciplinary Reviews: Computational Statistics},
  2(1):54--60, 2010.

\end{thebibliography}

\end{document}